\documentstyle[aaspp4,11pt]{article}
\catcode`\@=11
\def\bibstyle@aa{\bibpunct{(}{)}{;}{a}{}{,}}
\def\bibstyle@pass{\bibpunct{(}{)}{;}{a}{,}{,}}
\def\bibstyle@anngeo{\bibpunct{(}{)}{;}{a}{,}{,}}
\def\bibstyle@agsm{\bibpunct{(}{)}{,}{a}{}{,}\gdef\harvardand{\&}}
\def\bibstyle@kluwer{\bibpunct{(}{)}{,}{a}{}{,}\gdef\harvardand{\&}}
\def\bibstyle@dcu{\bibpunct{(}{)}{;}{a}{;}{,}\gdef\harvardand{and}}
\def\bibstyle@agu{\bibpunct{[}{]}{;}{a}{,}{,}}
\def\bibstyle@nlinproc{\bibpunct{(}{)}{;}{a}{,}{,}}
\def\bibstyle@cospar{\bibpunct{/}{/}{,}{n}{}{}%
     \gdef\@biblabel##1{##1.}}
\def\bibstyle@esa{\bibpunct{(}{)}{,}{n}{}{}%
     \gdef\@biblabel##1{##1.\hspace{1em}}%
     \gdef\@cite##1##2##3{\@citebegin Ref.~##1\if@tempswa,
          ##3\fi\@citeend}}
\def\bibstyle@nature{\bibpunct{}{}{,}{n}{}{}%
     \gdef\@biblabel##1{##1.}%
     \gdef\@cite##1##2##3{\unskip\mbox{$^{##1}$}}}
\def\bibstyle@plain{\bibpunct{[}{]}{,}{n}{}{}}
\let\bibstyle@alpha=\bibstyle@plain
\let\bibstyle@abbrv=\bibstyle@plain
\let\bibstyle@unsrt=\bibstyle@plain
\def\@cite#1#2#3{\if@tempswa\@citebegin\if#2\@empty\else#2 \fi
        #1\if#3\@empty\else, #3\fi\@citeend\else#1\fi}
\def\@citenum#1#2#3{\@citebegin\if@tempswa\if#2\@empty\else#2 \fi\fi
   #1\if#3\@empty\else, #3\fi\@citeend}
\def\@citexnum[#1][#2]#3{\if@filesw\immediate\write
      \@auxout{\string\citation{#3}}\fi
  \let\@citea\@empty
  \@cite{\@for\@citeb:=#3\do
    {\@citea\def\@citea{\@citesep\penalty\@m\ }%
     \def\@tempa##1##2\@nil{\edef\@citeb{\if##1\space##2\else##1##2\fi}}%
     \expandafter\@tempa\@citeb\@nil
     \@ifundefined{b@\@citeb}{%
       {\reset@font\bf ?}\@warning
       {Citation `\@citeb' on page \thepage \space undefined}}%
     \hbox{\csname b@\@citeb\endcsname}}}{#1}{#2}}
\def\@citex[#1][#2]#3{\let\@citea\@empty
  \@cite{\let\@citenm\@empty
    \@for\@citeb:=#3\do
    {\edef\@citeb{\expandafter\@iden\@citeb}%
     \if@filesw\immediate\write\@auxout{\string\citation{\@citeb}}\fi
     \@ifundefined{b@\@citeb}{\@citea%
       {\reset@font\bf ?}\@warning
       {Citation `\@citeb' on page \thepage \space undefined}}%
     {\let\@citemm=\@citenm
     \@cite@parse{\@citeb}%
     \if@tempswa
       \ifx\@citemm\@citenm\@yrsep\else\@citea{\@citenm}\@auyrsep\fi
       \ \@citedt \def\@citea{\@citesep\ }%
     \else
       \ifx\@citemm\@citenm, \@citedt\else\@citea{\@citenm}
           \@citebegin\@citedt\fi
       \def\@citea{\@citeend\@citesep\ }%
     \fi}}\if@tempswa\else\@citeend\fi}{#1}{#2}}
\def\@biblabel#1{\hfill}
\def\@biblabelnum#1{[#1]}
\def\@bibsetnum#1{\settowidth\labelwidth{\@biblabel{#1}}%
   \leftmargin\labelwidth \advance\leftmargin\labelsep
}
\def\@bibsetup#1{\leftmargin=1em\itemindent=-\leftmargin}
\def\@citebegin{(} \def\@citeend{)} \def\@citesep{;}
\def\@auyrsep{} \def\@yrsep{}
\def\bibstyle#1{\@ifundefined{bibstyle@#1}{\relax}
     {\csname bibstyle@#1\endcsname}}
\let\ori@document=\document
\def\document{\ori@document\global\let\bibstyle=\@gobble}
\def\bibpunct#1#2#3#4#5#6{\gdef\@citebegin{#1}\gdef\@citeend{#2}\gdef
   \@citesep{#3}\ifx #4n\global\let\@bibsetup=\@bibsetnum
   \global\let\@citex=\@citexnum
   \global\let\@biblabel=\@biblabelnum
   \global\let\@cite=\@citenum\fi
   \gdef\@auyrsep{#5}\gdef\@yrsep{#6}}
\newif\ifNAT@full\NAT@fullfalse
\def\cite{\@ifstar{\NAT@fulltrue\@citee}{\NAT@fullfalse\@citee}}
\def\@citee{\@ifnextchar [{\@tempswatrue\@citex@}{\@tempswafalse
    \@citex@[]}}
\def\@citex@[#1]{\@ifnextchar [{\@citex[#1]}{\@citex[][#1]}}
\newcommand{\citeauthor}[1]{\if@filesw\immediate\write
     \@auxout{\string\citation{#1}}\fi
\ifx\@citex\@citexnum
       {\reset@font\bf(author?)}\@warning
       {Cannot use \protect\citeauthor
         ^^J
        with numerical citations}\else
     \@ifundefined{b@#1}{%
       {\reset@font\bf ?}\@warning
       {Citation `#1' on page \thepage \space undefined}}%
       {\@cite@parse{#1}{\@citenm}}\fi}
\newcommand{\citeyear}[1]{\if@filesw\immediate\write
   \@auxout{\string\citation{#1}}\fi
\ifx\@citex\@citexnum
       {\reset@font\bf(year?)}\@warning
       {Cannot use \protect\citeyear
         ^^J
        with numerical citations}\else
     \@ifundefined{b@#1}{%
       {\reset@font\bf ?}\@warning
       {Citation `#1' on page \thepage \space undefined}}%
       {\@cite@parse{#1}\@citedt}\fi}
\newcommand{\citefullauthor}[1]{\if@filesw\immediate\write
     \@auxout{\string\citation{#1}}\fi
\ifx\@citex\@citexnum
       {\reset@font\bf(author?)}\@warning
       {Cannot use \protect\citeauthor
         ^^J
        with numerical citations}\else
     \@ifundefined{b@#1}{%
       {\reset@font\bf ?}\@warning
       {Citation `#1' on page \thepage \space undefined}}%
       {\@cite@parse{#1}{\@citefull}}\fi}
\def\@cite@parse#1{{%
      \@ifundefined{documentclass}
       {\let\prm=\relax\let\psf=\relax\let\ptt=\relax\let\pbf=\relax
        \let\psl=\relax\let\psc=\relax\let\pit=\relax\let\pem=\relax
        \let\prmfamily=\relax\let\psffamily=\relax\let\pttfamily=\relax
        \let\pbfseries=\relax\let\pslshape=\relax\let\pscshape=\relax
        \let\pitshape=\relax\let\pmdseries=\relax\let\pupshape=\relax
        \let\pc=\relax \let\pd=\relax \let\pb=\relax}
       {\let\protect=\@unexpandable@protect}%
     \xdef\@tempa{\csname b@#1\endcsname\relax}}%
     \expandafter\@citez\@tempa()()\@nil
     \ifNAT@full\let\@citenm\@citefull\fi}
\def\@citez#1(#2)#3()#4\@nil{\gdef\@citenm{#1}\gdef\@citedt{#2}%
  \ifx#3\relax\gdef\@citefull{#1}\else\gdef\@citefull{#3}\fi
  \if!#2!\expandafter\@citeapalk#1\@nil\fi}
\def\@citeapalk#1, #2\@nil{\gdef\@citenm{#1}\gdef\@citedt{#2}%
   \gdef\@citefull{#1}}
\newcommand{\citeauthoryear}[3]{\ifx#3\relax #1(#2)#1\else #2(#3)#1\fi}
\newcommand{\citestarts}{\@citebegin}
\newcommand{\citeends}{\@citeend}

\def\harvarditem{\@ifnextchar[{\@harvarditem}{\@harvarditem[\@empty]}}
\def\@harvarditem[#1]#2#3#4{\if!#1!\bibitem[#2(#3)]{#4}\else
  \bibitem[#1(#3)#2]{#4}\fi }
\newcommand{\harvardleft}{\@citebegin}
\newcommand{\harvardright}{\@citeend}
\newcommand{\harvardyearleft}{\@citebegin}
\newcommand{\harvardyearright}{\@citeend}
\def\harvardand{and}

\@ifundefined{chapter}{\def\bibsection{\section*{\refname
   \@mkboth{\uppercase{\refname}}{\uppercase{\refname}}}}}{\def
   \bibsection{\chapter*{\bibname
   \@mkboth{\uppercase{\bibname}}{\uppercase{\bibname}}}}}

\@ifundefined{reset@font}{\let\reset@font=\relax}{}

\catcode`\@=12
\newcommand\psr{PSR B1257+12}
\newcommand\bfreq{\mbox{\bf f}}
\newcommand\bk{\mbox{\bf k}}
\newcommand\sbk{\mbox{\scriptsize\bf k}}
\newcommand\eq[1]{(\ref{#1})}

\newcommand\etal{{\it et al.}}

\slugcomment{Submitted to {\em The Astrophysical Journal} on \today}

\lefthead{Konacki et al.}
\righthead{PSR B1257+12 Planetaty System}

\begin{document}
%\twocolumn
\title{Resonance in PSR B1257+12 Planetary System}

\author{Maciej Konacki,
 Andrzej J.Maciejewski and Alex Wolszczan  \altaffilmark{1}}
\affil{Toru\'n Centre for Astronomy, 
Nicolaus Copernicus University,\\
87-100 Toru\'n, Gagarina 11, Poland}

\altaffiltext{1}{Department of Astronomy and Astrophysics, The 
Pennsylvania State University}

\begin{abstract}
In this paper we present a new method that can be used for analysis of time 
of arrival  of a pulsar pulses (TOAs). It is designated especially to detect
quasi-periodic variations of TOAs. We apply our method to timing observations 
of PSR B1257+12 and demonstrate  that using it it is possible to detect  not only  
first harmonics of  a periodic variations, but also  the presence of a resonance 
effect. The resonance effect detected, independently of its  physical origin, can 
appear only when there is a non-linear interaction between two periodic modes. The 
explanation of TOAs variations as an effect of the existence of planets is, till now, 
the only known and well justified. In this context, the existence of the resonance 
frequency in TOAs is the most significant signature of the gravitational interaction 
of planets. 
\end{abstract}


\section{Introduction}

The first extra-solar planetary system was discovered by \cite{Wolszczan:92::}
around a millisecond radio pulsar, PSR B1257+12. The three
planets orbiting the pulsar have been
indirectly deduced from the analysis of quasi-periodic changes in the times
of arrival (TOAs) of pulses caused by the pulsar's reflex motion
around the center of mass of the system. In the analyses of this kind,
it is particularly important to establish a reliable method of distinguishing
planetary signatures from possible TOA variations of physically 
different origin.  
In the case of PSR B1257+12, it was possible to make this distinction 
and confirm the pulsar planets through the detection of mutual
gravitational perturbations between planets B and C \cite[]{Wolszczan:94::},
following predictions of the existence of this effect by \cite{Rasio:92::}, 
\cite{Malhotra:92::} (see also \cite{Malhotra:93::}, \cite{Rasio:93::} 
and \cite{Peale:93::}).

Practical methods of detection of the TOA variations caused by orbiting planets
include direct fits of Keplerian orbits to the TOA or TOA residual data
\cite[e.g.]{Thorsett:92::,Lazio:95::} and model--independent frequency domain approaches based on
Fourier transform techniques \cite[]{Konacki:96::,Bell:97::}. In fact, it 
appears that it is best to search for periodicities in TOAs (or post-fit TOA 
residuals left over from fits of the standard timing models) by examining
periodograms of the data, and then refine the search by fitting orbits
in the time domain using initial orbital parameters derived from a frequency 
domain analysis. 

The presence of planets around a pulsar 
causes pulse TOA variations which, for planets moving in
orbits with small eccentricities, have a quasi-periodic 
character and generate predictable, orbital element-dependent
features in the spectra of TOA residuals. This has led  
\cite[]{Konacki:96::} to devising a method of TOA residual analysis
based on the idea of a successive elimination of periodic terms applied
by Laskar (1992) in his frequency analysis of chaos in dynamical
systems. The frequency analysis provides
an efficient way to decompose a signal representing the TOA residual
variations into its harmonic components and study them 
in an entirely model-independent manner. As shown in
\cite[]{Konacki:96::} this method works perfectly well under
idealized conditions in which covariances among different parameters
of the timing model are negligible.

In this paper, we present an improved scheme for the frequency analysis of
pulsar timing observations in which a successive elimination of periodicities
in TOAs is incorporated in the modelling process rather than being applied
to the post-fit residuals. This makes the results obtained with our method
less sensitive to the effect of significant covariances which may exist
between various timing model parameters. We apply the frequency analysis
to TOA measurements of the planet pulsar, PSR B1257+12, using
a computer code developed to fit spectral timing models to data.
We show that our method allows an easy detection of the fundamental
orbital frequencies of planets A, B and C in the pulsar system and the
first harmonics of the frequencies of planets B and C generated by
eccentricities of the planetary orbits. Furthermore, by detecting
the effect of perturbations between planets B and C, we demonstrate
that the frequency analysis method represents a sensitive, model-independent
tool to analyze nonlinear interactions between periodic modes of processes of
various physical origins.

\section{Development of the method}

In the Fourier spectrum of a periodic process we detect the basic periodicity and some or all its harmonics. In the case of a quasi-periodic process, the Fourier spectrum contains a finite number of basic frequencies and their combinations with integer coefficients. A quasi-periodic signal represents a signature of the presence of interacting periodic processes. According to this description a real quasi-periodic function $S(t)$ of time $t$ can be written as a multiple Fourier series
\begin{equation}
\label{e:qp}
S(t) = \sum_{\sbk} s_{\sbk}\exp[2\pi {\rm i} \langle\bk ,\bfreq\rangle t] =
\sum_{\langle{\bf k ,\bf f}\rangle\geq 0} 
       a_{\sbk}\cos(2\pi\langle\bk ,\bfreq\rangle t)  + 
       b_{\sbk}\sin(2\pi\langle\bk ,\bfreq\rangle t),  
\end{equation}
where
\[
\langle\bk ,\bfreq\rangle = \sum_{l=1}^nk_lf_l, 
\] 
\[
\bfreq=(f_1, \ldots, f_n), \qquad \bk = (k_1,\ldots, k_n), 
\]
denote the vectors of basic frequencies, and the multi-index, respectively, and
$s_{\sbk}$, $a_{\sbk}$, $b_{\sbk}$ are complex and real amplitudes corresponding to the frequency $\langle\bk,\bfreq\rangle$.  In \eq{e:qp} the first sum is taken over all possible multi-indices $\bk$ with integer components $k_i$; the second sum is taken over all  multi-indices $\bk$ satisfying condition  
$\langle{\bf k ,\bf f}\rangle\geq 0$. A finite sum \eq{e:qp} models a quasi-periodic signal of arbitrary origin with the amplitude of the signal corresponding to  frequency 
$\langle\bk,\bfreq\rangle$ expressed as: 
\[
A_{\sbk}= \sqrt{ a_{\sbk}^2 + b_{\sbk}^2 } = 2 |s_{\sbk}|.
\]

Let us consider a time series of observations $\{\psi_i\}_{i=1}^N$, representing
a quasi-periodic process.
If the influence of noise and sampling effects is negligible, the spectrum of  $\{\psi_i\}_{i=1}^N$ will be characterized by a finite set of basic frequencies. 
In practice, any application of equation \eq{e:qp} to describe real data must 
properly account for the following facts: (a) data are unevenly sampled and
contaminated by noise, (b) a possible range of signal amplitudes
$[A_{min},A_{max}]$ can be very large, (c) periods corresponding to some
harmonics present in the signal may be longer than the data span.

We approach the above problem using the following algorithm. Let $R_0=\{r_l^{0}\}_{l=1}^N$ denote a set of residuals obtained by fitting the standard pulsar model $\varphi(t)$ to observations. 
Using the least squares method, we fit a function $F_{(1)}(t)= \varphi(t) + a_1  \cos(2\pi f_1 t)  + b_1 \sin(2\pi f_1 t)$ to the 
 observations  $\{\psi_l\}_{l=1}^N$. As the first approximation of $f_1$, 
 we take a frequency corresponding to the maximum in the periodogram 
of $R_0$. For this purpose we use  Lomb-Scargle periodogram \cite[]{Lomb:76::,Scargle:82::,Press:92::}. 

After the $k$-th iteration, we have assembled a set of residuals $ R_k=\{r_l^{(k)}\}_{l=1}^N$ defined as
\begin{equation}
\label{e:res}
r_l^{(k)}= \psi_l - F_{(k)}(t_l) , \qquad  l = 1, \ldots, N,
\end{equation}
where
\[
F_{(k)}(t) = F_{(k-1)}(t) + a_k \cos(2\pi f_k t) + 
b_k \sin(2 \pi f_k t), \quad F_{(0)}(t)\equiv 
\varphi(t),
\]
and 
$t_l$, $l = 1, \ldots, N$ denote observation times. 
Using the periodogram of $R_k$, we estimate $f_{k+1}$ and 
fit $F_{(k+1)}$ to the {\em original data} $\{\psi_i\}_{i=1}^N$. 
The whole process can be continued until a desired number of terms is obtained, or the final residuals fall below a predefined limit. 

This algorithm represents an {\em unconstrained frequency analysis},
because we have implicitly assumed that 
all frequencies present in the signal are independent. If fewer basic frequencies are present in the signal, our algorithm can be modified 
to a constrained form. First, we examine a predefined
number of basic frequencies $f_{1}, \ldots, f_{k}$ to obtain 
residuals $R_k$. Then, we {\em assume} that the dominant frequency $f_{k+1}$ of the process is a predefined combination of 
$f_{1}, \ldots, f_{k}$,
\begin{equation}
\label{e:lc}
f_{k+1}= l_1 f_{1} + \cdots +l_k f_k,
\end{equation}
where $l_j$, for $j=1,\ldots, k$ are integers and, as in the unconstrained case, we fit $F_{(k+1)}$ to the original data. Of course, in this case, $f_{k+1}$ {\em is not a free parameter}.  If necessary, this process can 
be repeated for frequencies $f_{k+c}$, $c=1,2, \ldots$.  In order 
to apply this algorithm, the coefficients of linear combinations \eq{e:lc} for all frequencies $f_{k+c}$, $c=1,2, \ldots$ 
have to be known in advance. In practice, it is conceivable that the signal 
amplitude at a constrained frequency \eq{e:lc} is greater than the amplitude at one or more basic frequencies. Consequently, an advance knowledge of the
ordering of frequencies with respect to corresponding signal amplitudes is
also necessary.
This ordering can be easily derived from the unconstrained 
frequency analysis. We call this version of our algorithm the {\em constrained frequency analysis}.

 Let us apply this approach to analyze TOAs of a pulsar with $N$ planets moving in 
orbits with small eccentricities. In the barycentric reference frame, with 
$z$-axis directed along the line of sight, changes in TOAs are proportional 
to the $z$-component of the pulsar radius vector.
As stated by
the KAM theorem \cite[]{Arnold:78::},
for realistic
planetary systems with weak interactions between planets,    
 coordinates and velocities of planets are almost always quasi-periodic 
functions of time. Consequently, we assume that the motions of pulsar
planets and hence the variations in a $z$-component of the pulsar vector
are quasi-periodic.
A presence of particular frequencies in the spectrum of TOAs is, in fact,
predictable, namely,  
 peaks with largest amplitudes should correspond to the basic (orbital) frequencies $f_1, \ldots, f_N$ of the planets. Of course, the hierarchy of amplitudes depends on masses and semi-major axes of the planets.  If eccentricities of orbits are greater than zero then 
 the first harmonics of frequencies $2f_1, \ldots, 2f_N$ will have significant amplitudes. Ratios of spectral amplitudes at the fundamental frequencies and their first harmonics depend on the eccentricities of planetary orbits.  If resonances are not present in a planetary system, 
 the amplitude corresponding to frequency $l_1f_1+\cdots+l_Nf_N$ decreases as $l=|l_1|+\cdots+|l_N|$ increases. 

A simple resonance in a planetary system exists, if orbital periods of two planets $p$ and $q$ in the system are commensurate. Approximate resonance relationships, which represent most of the practical cases, are also treated as resonances. Thus, there is a {\em simple orbital resonance} $l_p:l_q$  between two planets $p$ and $q$ if   frequency $f_R=l_pf_p + l_qf_q$ is small. In such a case,  amplitudes 
corresponding to frequencies $f^\pm_p=f_p \pm f_R$ and  $f^\pm_q=f_q \pm f_R$ are significant. The effect of resonance is caused by mutual interaction between 
planets, and thus its detecting in the spectra is a indirect confirmation of such interaction. Here we want to underline  that {\em the effect of the  resonance does not manifest itself in observations by the presence 
of a significant harmonic term with the resonance frequency}.      The most characteristic appearance of a simple  resonance is the presence of pairs of frequencies $f^\pm_q$ and $f^\pm_p$ located symmetrically around basic frequencies $f_p$ and $f_q$, respectively. Thus, {\em the total effect of a resonance is a sum of four harmonic terms with frequencies $f^\pm_q$ and $f^\pm_p$}.   
\section{Tests and Results}
0bservations of \psr\ used to test the analysis method outlined above consist of 217 daily averaged  TOAs covering the period in excess of
four years. As shown by \cite{Wolszczan:94::}, the observed TOA variations have a quasi-periodic character down to a $\sim$3 $\mu$s limit determined
by noise. 

The goal of our analysis was to identify all basic frequencies and their
harmonics in the signal and, possibly to detect the characteristic
signatures of a resonance effect.
\subsection{ Frequency Analysis}
Our test analysis has been performed as follows.
We generated fake TOAs for a pulsar orbited by three gravitationally  
interacting planets with orbital parameters chosen in such a way that the 
resonance had a period of 2059.2 days. These parameters were close to those given by \cite{Wolszczan:94::}. The simulation covered a period of about eight and 
a half years. For the first half of the period fake TOAs were generated 
at the times of real observations, whereas for the second half TOAs were generated at
the moments of real observations shifted by 4.22 years. A gaussian noise 
with the dispersion of 3$\mu$s was added to the synthetic TOAs.     

In the first test, we performed a constrained frequency analysis of both data sets choosing three independent frequencies $f_A$, $f_B$, and $f_C$ corresponding 
to the orbital periods of the respective planets. As shown in Fig. 1, five different frequencies (ordered here by decreasing amplitudes) were 
identified in both data sets: $f_1=f_C$, $f_2=f_B$, $f_3=2f_C$, $f_4=2f_B$, 
$f_5=f_A$. 

\placefigure{f:fa}

The unconstrained frequency analysis of the original and simulated TOAs detects peak 
frequencies listed in Table~\ref{t:two}. 
We have examined 
a hypothesis that the detected peak frequencies $f_3$ and $f_4$ are indeed the first harmonics
of $f_1$ and $f_2$ in the following way. 
Employing the frequency analysis, we fitted to data
a quasi-periodic model representing a sum of five harmonic terms at 
frequencies: $f_i$, $i=1,\ldots 5$. In this process, frequencies $f_3$ and 
$f_4$ had fixed values inside an interval close to twice the basic 
planetary frequencies of planets C and B, respectively. After the fit, 
we calculated the values of two parameters $\alpha_1 = f_3/f_1$ and $\alpha_2=f_4/f_2$ 
(so as $f_3 = \alpha_1 f_1$ and $f_4 = \alpha_2 f_2$) to obtain 
the weighted $\chi^2$ as a function of $(\alpha_1,\alpha_2)$. 
The results of this test (Fig.~\ref{f:chi}) demonstrate that, within
the limits set by a precision of the determination of frequencies   
$f_1$ and $f_2$ 
used to calculate $\alpha_1$ and $\alpha_2$, 
frequencies $f_3$ and $f_4$
are indeed the first harmonics of the fundamental frequencies of planets C and B.
\subsection{Detection of the resonance effect}
A direct frequency analysis of the original TOAs has failed to detect
the effect of resonance. Since the rms TOA residual is $\sim$3$\mu$s compared
to a predicted amplitude of the resonance
term of about 1$\mu$s, this result is not surprising.  
However, there are four 
harmonic terms related to the resonance and their sum obviously has an amplitude that is
higher than the noise level. Furthermore, the four frequencies related to the resonance lie symmetrically around the basic frequencies as is clearly visible 
in  Fig.~\ref{f:ideal}. With this knowledge, and a fixed value of $f_R$, we fited to data a
quasi-periodic model with frequencies: $f_1$, $f_2$, $f_3=2f_1$, $f_4=2f_2$, 
$f_5$, $f_{6,7}=f_1 \pm f_R$, $f_{8,9}=f_2\pm f_R$. Varying $f_R$
we minimizeded a weighed $\chi^2$ as a function of $f_R$
to isolate any distinguished frequency $f_R$, whose presence might
indicate the existence of a resonance effect between the observed TOA periodicities.

The result of this test for real data is shown in Fig.~\ref{f:rfr}. 
Although
a position of the minimum, $f=6.9\times10^{-4}$, does not coincide with that
predicted by the theory, it is well defined indicating the
detection of a real effect. The observed displacement of a $\chi^2$ minimum
can be easily explained by the effect of a limited data span as shown in
Fig. 5 which demonstrates that,
with an increasing time coverage, the resonance frequency is determined
with an increasing precision.
Finally, 
to validate this test further, we repeated it with the data 
consisting of artificial TOAs
modulated by Keplerian (i.e. noninteracting) planetary orbits. Clearly, 
in this case the minimum occurs at $f_R = 0$, as expected in the absence
of any resonance effect. 
Consequently, we confirm the presence of a resonance effect in the TOAs
for PSR B1257+12, as originally demonstrated by Wolszczan (1994) using
a time domain, real orbit analysis.

\subsection{Keplerian elements of planets}

Given the values of fundamental frequencies and amplitudes of
their harmonics, orbital elements of planets can be calculated,
assuming that they move in Keplerian orbits \cite[]{Konacki:96::}.
Planetary masses can be calculated by assuming
a pulsar mass  to be $M_{\star} = 1.4M_{\odot}$) and by setting
orbital inclinations to 90$^{\circ}$.
We find that $e = 0.0$, $m = 0.0168\,M_{\oplus}$, $P = 2189095$~(s) for planet A (only the fundamental frequency is significant so we assume that the planet's 
eccentricity is zero), $e = 0.0178$, $m = 3.40\,M_{\oplus}$, $P = 5748596$~(s) 
for planet B and $e = 0.0229$, $m = 2.83\,M_{\oplus}$, $P = 8486539$~(s) for planet 
C. Finding $\omega$ and $T_p$ is also possible but it is more complicated.  
Clearly, these values are somewhat different from those given in
Table~\ref{t:one}(especially eccentricities). This is understandable, given  
that the frequency analysis-based model (constrained or unconstrained) 
and the one based on Keplerian orbits are not entirely equivalent. In the case of 
Keplerian orbits, frequencies and amplitudes are not independent parameters, 
while in the case 
of constrained frequency analysis only frequencies are constrained (as integer
combinations of some fundamental frequencies) with all amplitudes remaining 
independent.

Clearly, the frequency analysis is not particularly convenient to determine
Keplerian elements of planets.
It is better to simply fit a suitable number of Keplerian orbits
to observations. However, the frequency analysis procedure provides a good test
of consistency with the results of fitting
Keplerian orbits in time domain. It is encouraging to see that PSR B1257+12 planetary system does pass this test.

\acknowledgments{MK and AJM were supported by the KBN grant 2P03D.023.10. AW
was supported from the NASA grant NAG5-4301 and the NSF grant AST-9619552.

%\bibliographystyle{apj}
%\bibliography{apj,books,psr}
\newcommand{\noopsort}[1]{}

\begin{deluxetable}{lrrr}
\tablewidth{27em}
\tablecaption{PSR B1257+12 Planets Orbital Parameters \label{t:one}}
\tablehead{
\colhead{Parameter}               & \colhead{A} &
\colhead{B} & \colhead{C}    }

\startdata

$a \sin i$ (ms)         & 0.0035(6)       & 1.3106(6)       & 1.4121(6) \nl
$e$                     &  0.0            & 0.0182(9)       & 0.0264(9) \nl
$T_0$ (JD)              &  2448754.3(7)   & 2448770.3(6)    & 2448784.4(6) \nl
$P$  (s)                &  2189645(4000)  & 5748713(90)     & 8486447(180) \nl
$f$ (deg)          & 0.0             & 249(3)          & 106(2)       \nl
$m$ ($\mbox{M}_\oplus$) & $0.015/\sin i_A$& $3.4/\sin i_B$& $2.8/\sin i_C$  \nl
$r$ (AU)                & 0.19            & 0.36            & 0.47        \nl
\hline
\enddata
\end{deluxetable}

\begin{deluxetable}{lrr}
\tablewidth{20em}
\tablecaption{Frequencies found in real and simulated observations\label{t:two}}\tablehead{
\colhead{$f$}               & \colhead{Real TOAs} &
\colhead{Simulation}  }
\startdata

  $f_1$        & 0.01018085      &0.01001988  \nl
  $f_2$        & 0.01502974      &0.01478685  \nl
  $f_3$        & 0.02035317      &0.02003982  \nl
  $f_4$        & 0.03004729      &0.02957544  \nl
  $f_5$        & 0.03947914      &0.03943777  \nl
$2f_1-f_3$      & 8.53048$\cdot10^{-6}$ &-5.44713$\cdot10^{-6}$\nl
$2f_2-f_4$      & 1.21856$\cdot10^{-5}$ &-1.72562$\cdot10^{-6}$\nl
\hline
\enddata
\end{deluxetable}

\clearpage
\newpage
\figcaption[fa.eps]{
Frequency analysis of the real (left hand side) and simulated (right hand side)  
observations of \psr . The top graphs (a and b) show TOA residuals for the real (a) and simulated (b) data after fitting the standard timing model. All other 
pictures are Lomb-Scargle normalized periodograms (LSNP) for which on the $x$-axis 
we have frequencies (in 1/day) and on the $y$-axis we have power units of LSNP 
(Press \etal, 1992, p.577). A number near a pick in each periodogram is equal 
to the amplitude (in microseconds) of the component of the signal. In the second 
row, in LSNP of residuals $R_0$ for the real (c) and simulated (d) data we can see 
only fundamental frequencies of planets B and C. In the next step, in LSNP of $R_2$ 
we can see frequencies of second harmonics of planets B and C for both data 
sets (pictures e and f). Having subtracted them all from the signal (now, by 
fitting $F_4(t)$ to the data) we can see the fundamental frequency of the planet 
A (g and h). And finally, after additionaly subtracting the fundamental term of the planet A (by fitting $F_5(t)$ to the data), we see that in residuals $R_5$ 
of the real data (i) there is not a dominant pick, however one can see at least one of the peaks corresponding to the resonance frequencies $f_B \pm f_R$ and  
$f_C \pm f_R$ in LSNP of the fake data (j). In an idealized situation when
the data densely cover the time domain, the respective periodogram looks like 
in Fig.~2.
\label{f:fa}
}
\figcaption[ideal.eps]{
LSNP of the last step of the frequency analysis of an idealized data set 
(one TOA per day without noise for the time span of two resonance periods). 
This figure corresponds to the pictures (i) and (j) in Fig.~1.\label{f:ideal}
}
\figcaption[chi.ps]{
Test for the existence of the second harmonics of the planet fundamental 
frequencies. In the picture, $\alpha_1$ is the ratio of the fundamental 
frequency to the frequency $f_3$ (suspected to be the second harmonic) for 
the planet C and $\alpha_2$ is the corresponding value for the planet B. 
Contours correspond to confidence ellipses with marked confidence levels
for $\Delta\chi^2(\alpha_1,\alpha_2) = \chi^2(\alpha_1,\alpha_2) - \chi^2_0$
where $\chi^2_0$ is the global minimum. 
\label{f:chi}
}
\figcaption[rfr.eps]{
Detecting the resonance effect in the real data. In the picture, $f_R$ is 
a frequency in $1/$day. Confidence levels $1\sigma$, $2\sigma$, $3\sigma$ 
for $\Delta\chi^2(f_R) = \chi^2(f_R) - \chi^2_0$ where $\chi^2_0$ is the 
global minimum are shown. The minimum is well established around the frequency 
$f=6.9\times10^{-3}$ that is shifted with respect to the resonance frequency 
$f_r=4.86\times 10^{-3}$ predicted by the theory. It results directly from 
the feature of the real data covering the time span shorter then the period 
of the resonance.
\label{f:rfr}
}
\figcaption[all.eps]{ 
Detecting the resonace effect in the simulated TOAs of various origin.
In the picture, $f_R$ is a frequency in $1/$day and $\Delta\chi^2(f_R)$
is calculated as a departure from the global minimumu found for each case.
(a) Test for the TOAs modulated by Keplerian planetary model. (b,c,d) 
Detecting the resonance for the fake TOAs resembling real observations but 
computed for different times spans: (b) data cover 0.75 of the resonance
period, (c) data cover 1.0 of the resonance period, (d) data cover 2.0 of
the resonance period. Increasing the time span allows to find the proper
value of the resonance frequency.
\label{f:all}
}

%
%---------1
\begin{figure}
\figurenum{1}
\epsscale{0.8}
\plotone{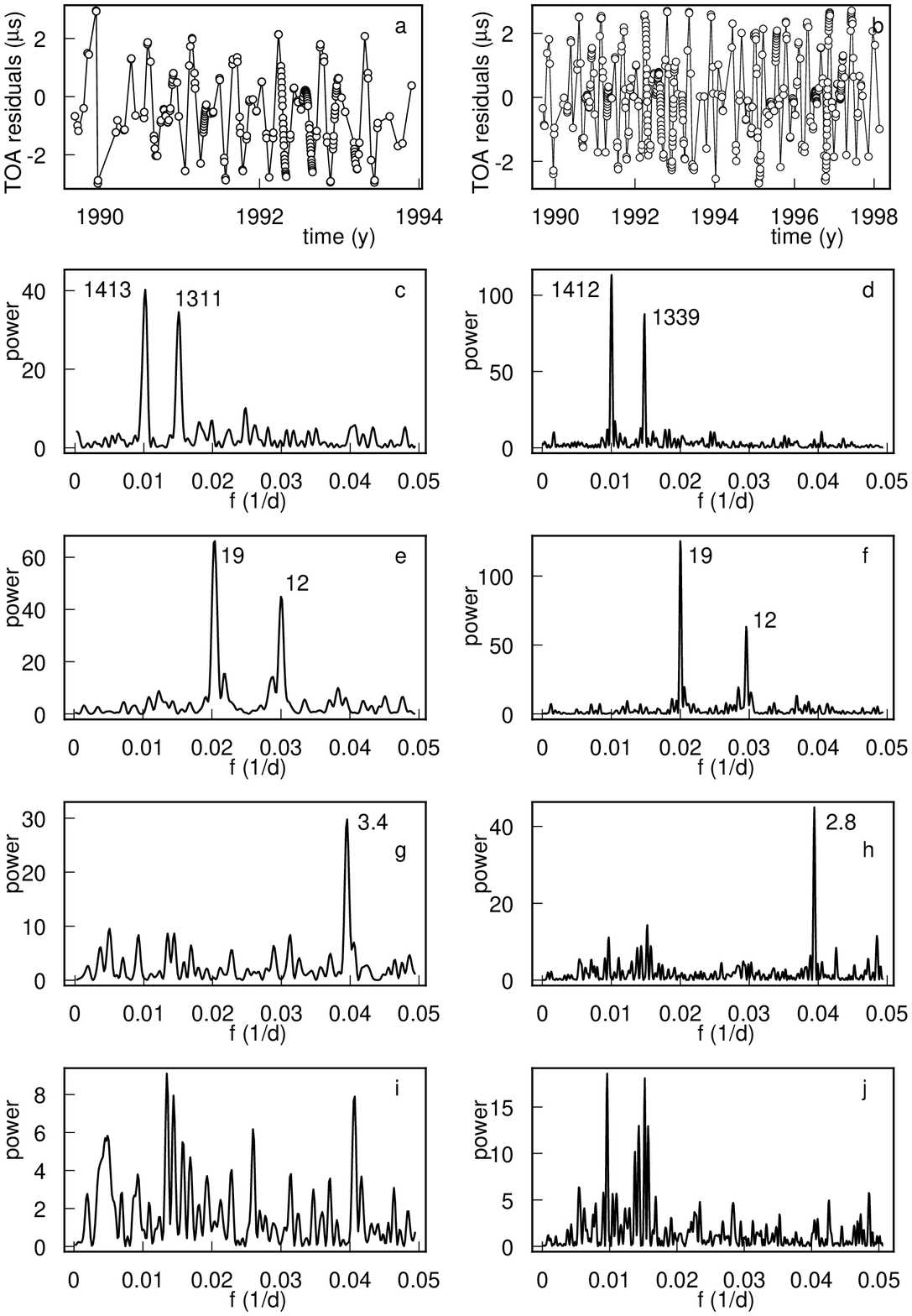}
\caption{}
\end{figure}
%
%---------2
\begin{figure}
\figurenum{2}
\epsscale{0.8}
\plotone{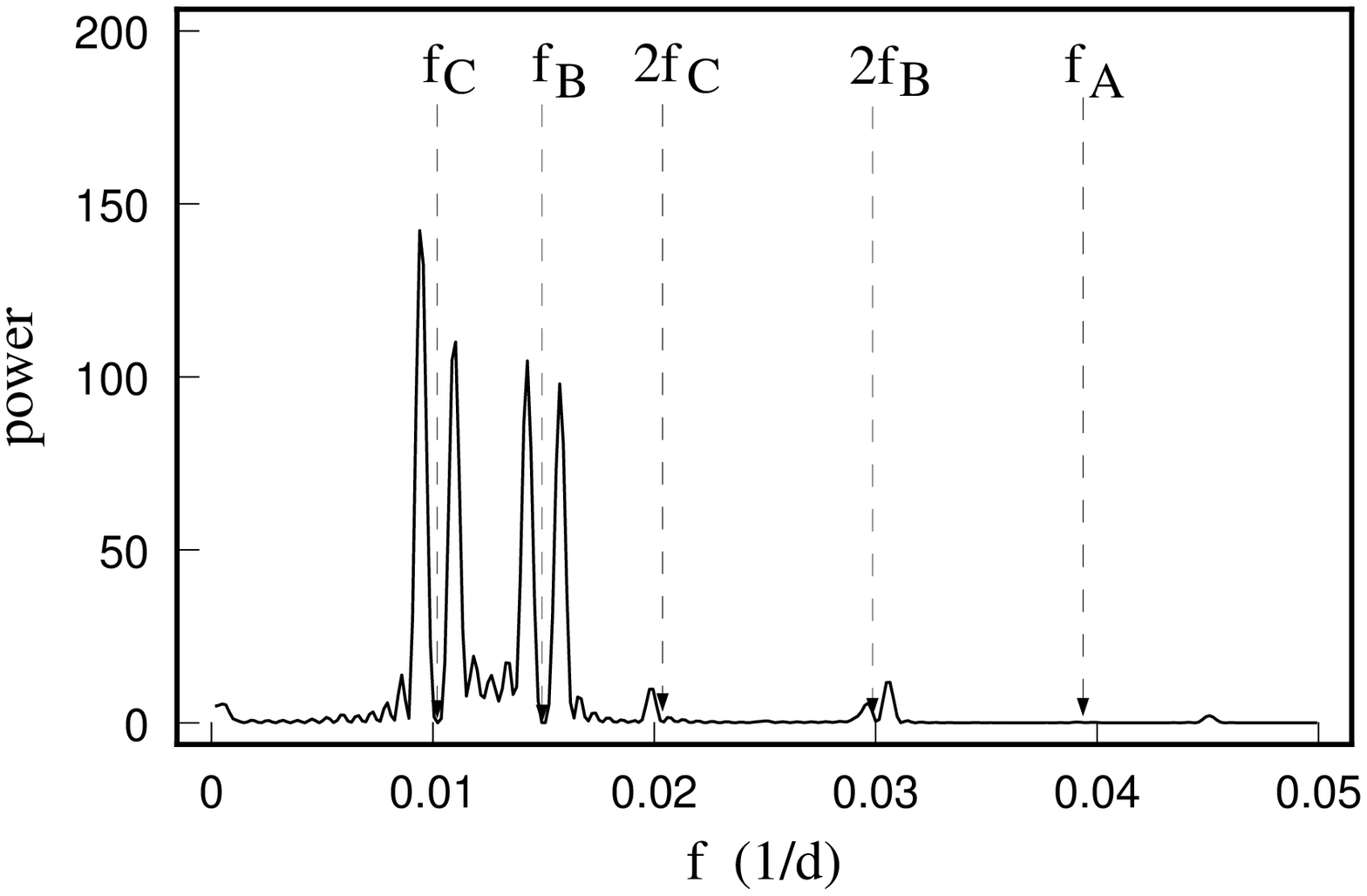}
\caption{}
\end{figure}
%
%---------3
\begin{figure}
\figurenum{3}
\epsscale{0.8}
\plotone{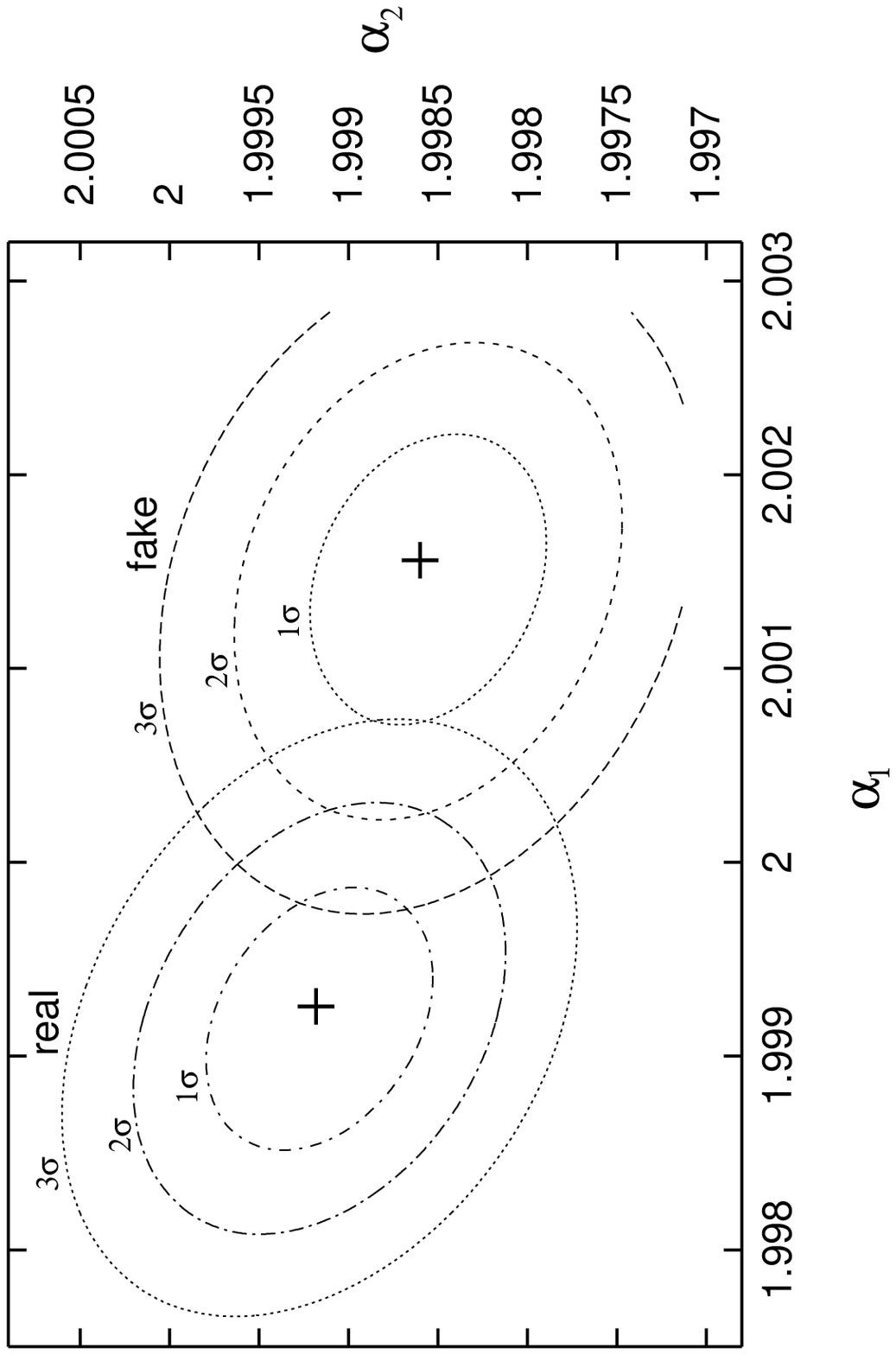}
\caption{}
\end{figure}
%
%---------4
\begin{figure}
\figurenum{4}
\epsscale{0.8}
\plotone{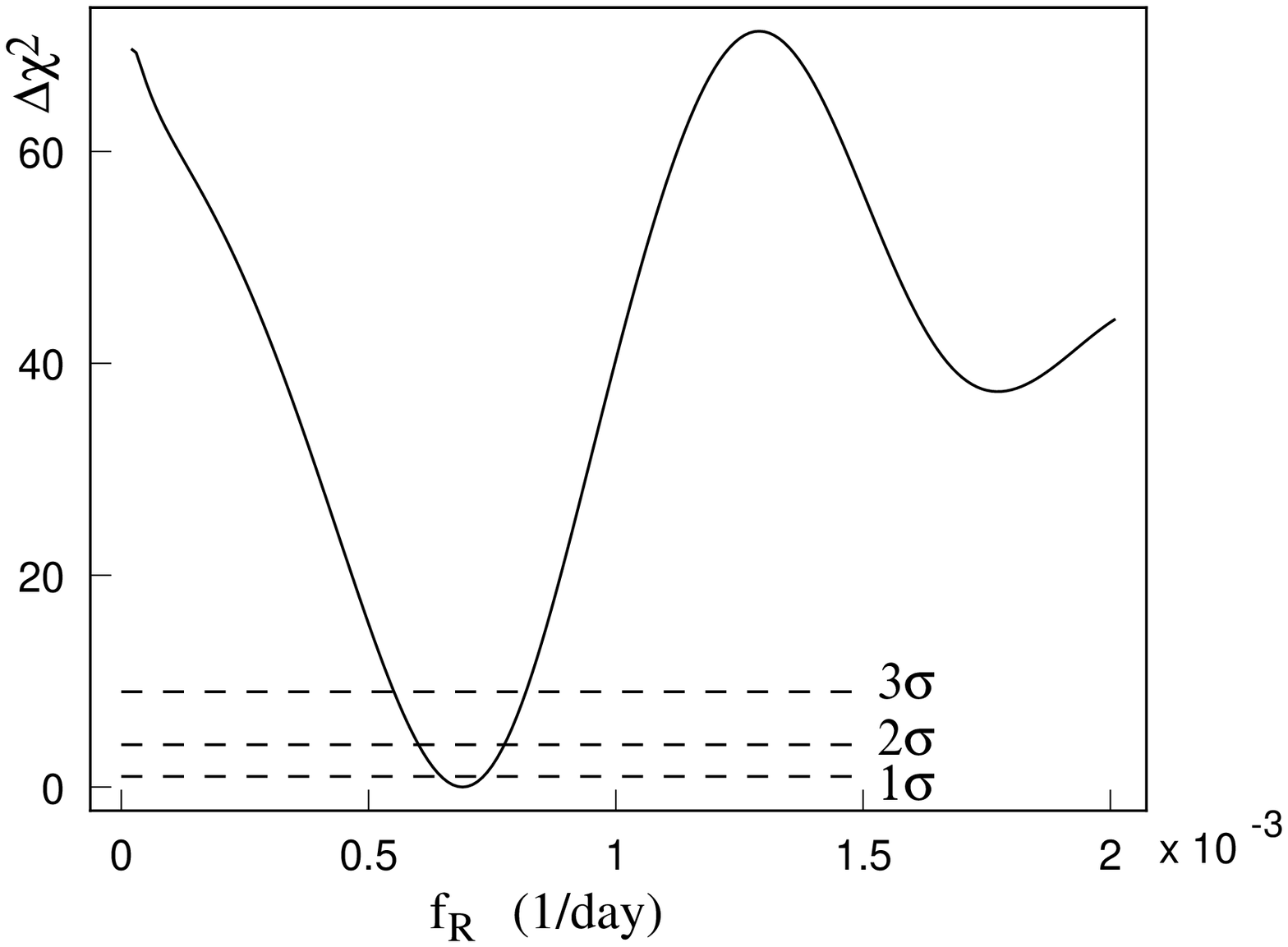}
\caption{}
\end{figure}
%
%---------5
\begin{figure}
\figurenum{5}
\epsscale{0.8}
\plotone{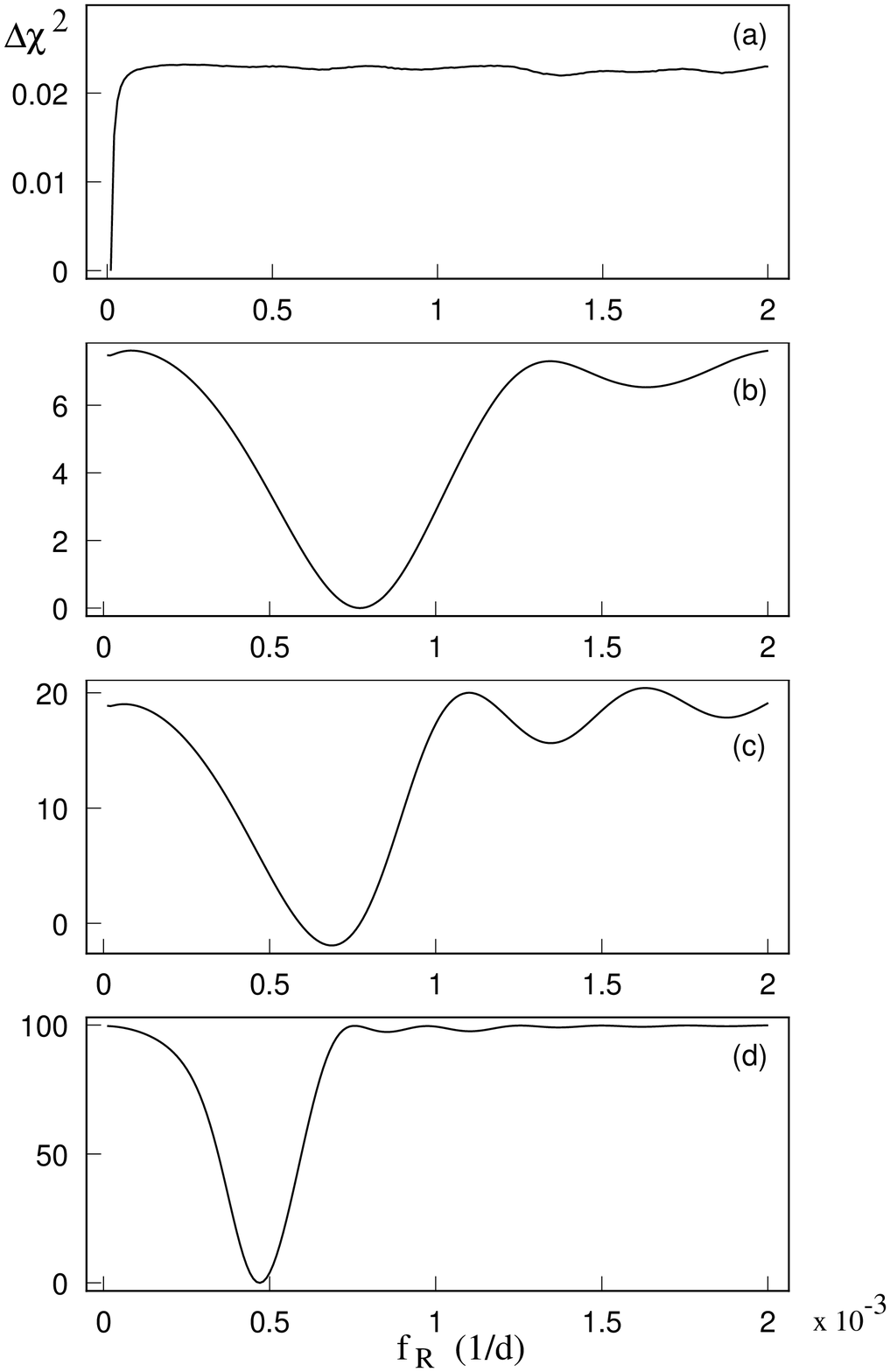}
\caption{}
\end{figure}

\end{document}